\pdfoutput=1
\documentclass[pre,aps,twocolumn,showpacs,floatfix,nofootinbib,
superscriptaddress
]{revtex4-1}
\usepackage{subfigure}
\usepackage{amssymb}
\usepackage{amsfonts}
\usepackage{amsmath}
\usepackage{amsthm}
\usepackage{epsfig}
\usepackage{graphicx}
\usepackage[usenames,dvipsnames]{color}
\usepackage[latin1]{inputenc}

\usepackage[hidelinks,unicode=true]{hyperref}
\hypersetup{colorlinks=true,
 	linkcolor=blue,
 	urlcolor=blue,
 	citecolor=blue,
 	pdfhighlight=/N
 }
 \usepackage{comment}
 \usepackage[normalem]{ulem}

\newcommand{\lrangle}[1]{\langle{#1}\rangle}

\newcommand{\alphaloc}[1]{\alpha^{_\text{#1}}_\text{loc}}
\begin{document}

\title{Local roughness exponent in the nonlinear molecular-beam-epitaxy universality class in one-dimension}

\author{Edwin E. Mozo Luis}
\email{eluis@ufba.br}
\address{Instituto de F\'{\i}sica, Universidade Federal da Bahia,
  Campus Universit\'{a}rio da Federa\c c\~ao,
  Rua Bar\~{a}o de Jeremoabo s/n, 40170-115, Salvador, BA, Brazil}
\author{Thiago A. de Assis}
\email{thiagoaa@ufba.br}
\address{Instituto de F\'{\i}sica, Universidade Federal da Bahia,
  Campus Universit\'{a}rio da Federa\c c\~ao,
  Rua Bar\~{a}o de Jeremoabo s/n, 40170-115, Salvador, BA, Brazil}
\author{Silvio C. Ferreira}
\email{silviojr@ufv.br}
\affiliation{Departamento de F\'{\i}sica, Universidade Federal de
 Vi\c{c}osa, Minas Gerais, 36570-900, Vi\c{c}osa, Brazil}
\affiliation{National Institute of Science and Technology for Complex Systems, 22290-180, Rio de Janeiro, Brazil}
 \author{Roberto F. S. Andrade}
\email{randrade@ufba.br}
\address{Instituto de F\'{\i}sica, Universidade Federal da Bahia,
	Campus Universit\'{a}rio da Federa\c c\~ao,
	Rua Bar\~{a}o de Jeremoabo s/n, 40170-115, Salvador, BA, Brazil}
\affiliation{National Institute of Science and Technology for Complex Systems, 22290-180, Rio de Janeiro, Brazil}

\begin{abstract}
We report local roughness exponents, $\alpha_{\text{loc}}$, for three interface growth models in one
dimension which are believed to belong the non-linear molecular-beam-epitaxy (nMBE)
universality class represented by the Villain-Lais-Das Sarma (VLDS) stochastic
equation. We applied an optimum detrended fluctuation analysis (ODFA) [Luis
\textit{et al.}, \href{http://dx.doi.org/ 10.1103/PhysRevE.95.042801}{Phys. Rev.
	E \textbf{95}, 042801 (2017)}] and compared the outcomes with standard
detrending methods. We observe in all investigated models that ODFA outperforms
the standard methods providing exponents in the narrow interval
$\alphaloc{}\in[0.96,0.98]$ consistent with renormalization group predictions
for the VLDS equation. In particular, these exponent values are calculated for the
Clarke-Vvdensky  and Das Sarma-Tamborenea models characterized by very
strong corrections to the scaling, for which large deviations of these values
had been reported. Our results strongly support the absence of anomalous
scaling in the nMBE universality class and the existence of corrections in the
form $\alphaloc{}=1-\epsilon$ of the one-loop
renormalization group analysis of the VLDS equation.

\end{abstract}

\maketitle

\section{Introduction}
\label{Int}

Kinetic roughening is an important feature related to the growth of interfaces
under nonequilibrium conditions~\cite{Barabasi1995,Krug1997}. In many systems
under specified conditions, including that of molecular-beam epitaxy (MBE)
experiments, the surface diffusion may be a ruling mechanism competing with the
deposition~\cite{Barabasi1995,Evans2006}. Stochastic modeling of MBE is a
frontline scientific issue since it corresponds to a technique to produce high
quality thin films for many applications~\cite{Evans2006,Ohring2002}. In the
simplest cases, the modeling assumes a limited mobility of adatoms. Some
examples include the conservative restricted solid-on-solid
(CRSOS)~\cite{Kim1994,Kim1997} and the Das Sarma and Tamborenea (DT)~\cite{DasSarma1991}
models, in which short-range surface diffusion and permanent aggregation take
place after adsorption. More realistic models include thermally activated
processes where the mobility is not limited. A noteworthy one is the
Clarke-Vvedensky (CV)
model~\cite{Clarke1987,Clarke1988,Kotrla1996,DeAssis2015b}, in which the adatom
diffusion rates follow Arrhenius laws, with energy barriers depending on the
local number of bonds. Recently, the scaling properties of a limited-mobility
model were compared with the CV model~\cite{To2018}, discussing the effects of
memory (non-Markovianity) and probabilities of adatom detachment from terrace
steps. It was observed that many central features of thermally activated models
can be captured with their limited mobility versions~\cite{To2018}.

The CRSOS~\cite{Park2001}, DT~\cite{DasSarma1991,Predota1996} and
CV~\cite{Clarke1987,Clarke1988,DeAssis2015b} models are connected with the
nonlinear molecular-beam-epitaxy (nMBE) universality class, since the surface
dynamic is ruled by adatom diffusion. If the incoming particle flux is omitted, the corresponding nMBE growth equation, also called
Vilain-Lai-Das Sarma (VLDS)~\cite{Villain1991,Lai1991}, is given by
\begin{equation}
\frac{\partial h}{\partial t} = -\nu_{4} \nabla^{4}h + \lambda_{4} \nabla^2\left(\nabla h\right)^{2} + \eta(\mathbf{x},t),
\label{VLDSeq}
\end{equation}
where $h$ corresponds to the height, at the position $\mathbf{x}$ and time $t$,
with respect to the initial $d$-dimensional substrate, $\nu_4$ and $\lambda_4$
are constants and $\eta(\mathbf{x},t)$ is a nonconservative Gaussian noise.

In this work, we investigate interface growth on one-dimensional substrates. The
growth ($\beta$) and the dynamical ($z$) exponents are used as benchmarks to
describe the interface scale invariance \cite{Barabasi1995}. The former exponent characterize how height fluctuations
$\omega$ while the latter how the characteristic correlation length $\xi$
evolve, usually  obeying scaling laws of the forms $\omega \sim t^\beta$ and
$\xi\sim t^{1/z}$, respectively. The global roughness exponent $\alpha=\beta z$ can
also be used in the regime where the correlation length is much larger than the
scale of observation $L$ when $\omega\sim L^\alpha$~\cite{Family1985,Barabasi1995}.
According to a dynamical one-loop renormalization-group (RG) analysis of
Eq.~(\ref{VLDSeq}), the roughness and dynamic exponents are given by $\alpha =
(4-d)/3$ and $z=(8+d)/3$. However, Janssen~\cite{Janssen1997} pointed out that
this conclusion was derived from an ill-defined transformation and there would
be higher order corrections. For instance, small
negative corrections to $\alpha$ and $z$ were reported in all
dimensions from a two-loop calculation ~\cite{Janssen1997}. These corrections are supported by numerical
results for CRSOS model~\cite{AaraoReis2004b}, in which $\alpha=0.94(2)<1$, the predicted value by the one-loop RG analysis $\alpha=1$.

Nonetheless, several investigations of nMBE lattice models formerly suggested that the local and
global height fluctuations scale with different local and global roughness exponents,
characterizing an anomalous scaling~\cite{Lopez2005,Lopez1999}. To define this
phenomenon, let us consider the interface fluctuations within a window of length
$r$ and at time $t$ (hereafter called quadratic local roughness)
\begin{equation}
\omega_i^2(r,t) ={\lrangle{h^2}_i-\lrangle{h}_i^2},
\label{eq:localrou}
\end{equation}
where $\lrangle{\cdots}_i$ means averages over the window $i$. The quadratic
local interface roughness $\omega^2(r,t)$ is defined considering the average of
$\omega_i^2$ over different windows and independent realizations. In normal
dynamical scaling, in which the Family-Vicsek ansatz \cite{Family1985} holds, the
local roughness for a window of length $r$ increases as $\omega\sim t^\beta$ for $t\ll r^z$ and saturates
as $\omega\sim r^{\alphaloc{}}$ for $t\gg r^z$, with $z=\alphaloc{}/\beta$. The
 exponent $\alphaloc{}$ is the local roughness or Hurst
exponent~\cite{Barabasi1995}.
{Anomalous scaling happens when local and global roughness exponents are
	different implying that the local roughness
	presents dependence on both window size and time at short scales given
	by $\omega(r,t)\sim r^{\alphaloc{}} t^{\kappa}$ with $\kappa=(\alpha-\alphaloc{})/z$.}

Anomalous scaling was reported in theoretical analyzes of epitaxial surface growth and numerical integration of the VLDS equation in $d =1$ and $2$ ~\cite{Xia2013}. Mound formation was claimed to justify the anomalous scaling, which contrasts with the conclusions reported in Refs.~\cite{Lopez2005,AaraoReis2013a}, according to which normal scaling should occur in local growth processes. Crystalline mounds have been also used to justify anomalous scaling in experiments~\cite{Yim2006}, while the interplay
between nonlocal strain and substrate disorder was pointed out as a mechanism involved in the anomalous scaling
in epitaxial growth of semiconductor CdTe films ~\cite{Nascimento2011,Mata2008}.

In the context of lattice models, it was reported an apparent anomalous scaling at short times which
asymptotically turns into normal scaling for the CRSOS
model~\cite{AaraoReis2013a}. The local roughness exponent of the DT model was
reported to be $\alphaloc{} \approx 0.7$ in $d=1$~\cite{Krug1994,Chame2004}.
This value is different from the global roughness exponent $\alpha=1-\epsilon$ predicted
in the RG analysis of the VLDS equation, which might suggest anomalous
roughening~\cite{Dasgupta1996,DasSarma1996}. However, the local roughness
distributions~\cite{AaraoReis2013a} suggest that the one-dimensional DT model has
normal scaling, in agreement with the predictions of dynamic RG analysis for
local growth process without quenched disorder or additional
symmetries~\cite{Lopez2005}. Thus, the DT model remains controversial and a careful
consideration regarding their local roughness exponents, especially
without noise reduction techniques~\cite{Punyindu1998,Xun2015,Disrattakit2016}, is
worthwhile.

An evaluation of $\alphaloc{}$ for CV model in two-dimensions, in agreement with
the nMBE class, was recently reported~\cite{Luis2017}. The effective roughness
exponent was calculated with the optimal detrended fluctuation analysis (ODFA),
while the exponents obtained with other methods did not match with those of the nMBE
class~\cite{Luis2017}. This result provided support for
non-anomalous asymptotic scaling in CV model, corroborating the claim that this
transient effect is a consequence of large corrections to the asymptotic normal
scaling~\cite{DeAssis2015b}. However, an explicit observation of the local roughness
exponent for the CV model in $d=1$ is still missing.

Motivated by the aforementioned studies, we present an analysis of the local
roughness exponent for CRSOS, DT and CV models in one-dimension using the ODFA
method. Our results show that the second order ODFA method suitably yields
values of $\alphaloc{}$ consistent with the nMBE class for CRSOS and CV models.
Moreover, the obtained exponents corroborate the existence of corrections in
the one-loop RG~\cite{Janssen1997}.  For DT models, the ODFA method also provides
values in agreement with nMBE class specially in the case of mild noise
reductions~\cite{Punyindu1998,Xun2015,Disrattakit2016}. In the original DT, two scaling regimes were observed: at short scales we report
$\alphaloc{(DT)}\approx 0.90$ and at intermediary ones $\alphaloc{(DT)}\approx
0.97$, both considerably closer to the nMBE class than the values found with
other methods. The results presented here are consistent with
the conjecture of Ref.~\cite{Lopez2005}, which argues that intrinsic
anomalous roughening cannot occur in local growth models.

The rest of this paper is organized as follows. In section~\ref{sec:modelbc}, we
present the models and basic concepts involved in this work. In
section~\ref{sec:sorderdfa}, we discuss the limits where the ODFA method
outperforms other methods, considering mounded initial conditions. In
section~\ref{sec:Slocal}, the scaling of surface roughness is analyzed
and the local roughness exponent is reported. In section~\ref{sec:conclu}, we
summarize our conclusions and prospects.

\section{Models and basic concepts}
\label{sec:modelbc}

The lattice models studied in this work are the CRSOS, DT and CV. All
simulations were performed on a  initially flat one-dimensional substrate with
$L$ sites and periodic boundary conditions. One time unit corresponds to the
deposition of $L$ adatoms. In all models, deposition occurs with rate $F=1$ in a
flux normal to the substrate and obeys a solid-on-solid
condition~\cite{Barabasi1995}.

In the CRSOS model, a site is randomly chosen for one
adatom deposition. The height differences $\delta h$ between nearest-neighbors
obey the restriction $\delta h \leqslant \delta H_{max}$. We consider the
case $\delta H_{max}=1$. If this condition is satisfied for the randomly chosen
incidence site, the particle permanently sticks there. Otherwise, it searches
the nearest position where the condition is satisfied, which becomes the place of deposition.
In the case of multiple options, one of them is randomly chosen.

In the CV model, deposition occurs at a constant and uniform rate while the
adatom diffusion rate is given by an Arrhenius law in the form $D=\nu_0
\exp(-E/k_{B}T)$ where $\nu_0$ is an attempt frequency, $k_B$ the Boltzmann
constant, and $E$ is an energy barrier for the hopping, which includes the
contribution of the substrate ($E_S$) and lateral bonds ($E_N$) assuming the
form $E=E_S+nE_N$. The ratio $R=D_{0}/F$, in which $D = D_{0} \varepsilon^{n}$ is
the hopping rate if an adatom has $n$ lateral neighbors, is a control parameter
of the model \cite{DeAssis2015b,Luis2017}. In this work we use $R=10$ and
$\varepsilon=0.01$, which leads to a large surface roughness, since it corresponds
to a low temperature (low mobility) regime.

In the one-dimensional DT model~\cite{DasSarma1991}, the arriving particle sticks at the top of the
incidence site if there is one or two lateral bonds. Otherwise, if one of
the nearest-neighbors satisfies this condition, it is chosen for the deposition
whereas if both do, one of them is randomly chosen. If neither the deposition
site nor any of the nearest-neighbors have lateral bonds,  the particle
sticks at the top of the incidence site. We also applied the noise
reduction technique~\cite{Xun2015}, in which a site must be selected $M$ times
before implementing a deposition. We considered the case of mild noise reduction
$M=4$ where the  interface roughness remains large.

A characteristic lateral surface
length can be estimated as the first zero ($\xi_{0}$) of the height-height
correlation function defined as~\cite{Evans2006,Leal2011a,Leal2011}
\begin{equation}
\Gamma(s,t) \equiv  \left\langle\tilde{h}( s_0+s,t)
\tilde{h}(s_0,t) \right \rangle,
\end{equation}
where $\tilde{h}\equiv h-\overline{h}$, and the averages are taken over
different initial positions $s_0$ and different configurations. The correlation
length $\xi_{0}$ is defined as the position of the first zero of the correlation
function, i.e. $\Gamma(\xi_0,t)=0$ and are expected to scale as $\xi_{0}(t) \sim
t^{1/z_c}$, where $z_c$ is the coarsening exponent that usually corresponds to the dynamical exponent defined previously.

\begin{figure*}[hbt!]
	\includegraphics [width=0.4936\linewidth] {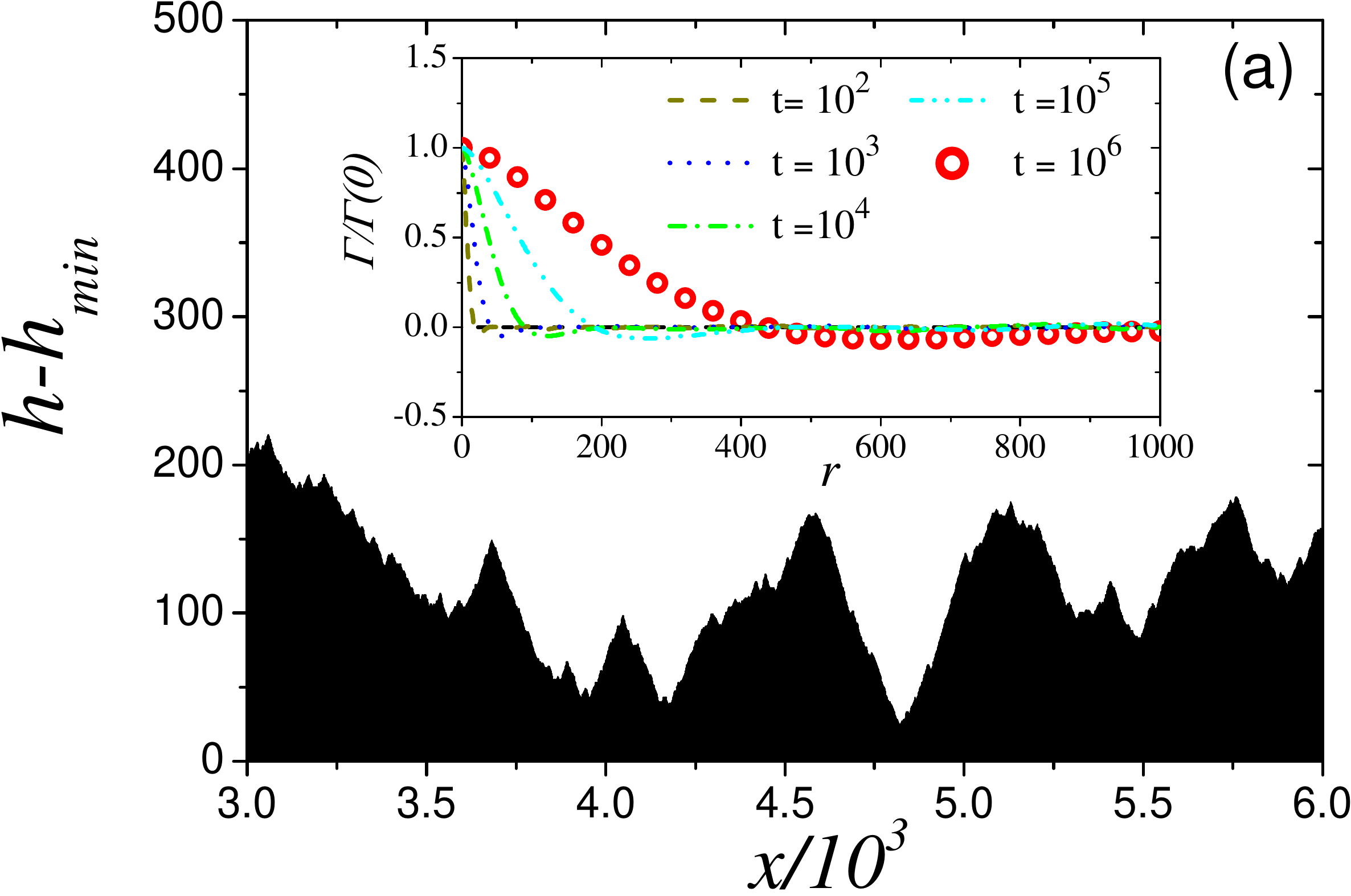}
	\includegraphics [width=0.4936\linewidth] {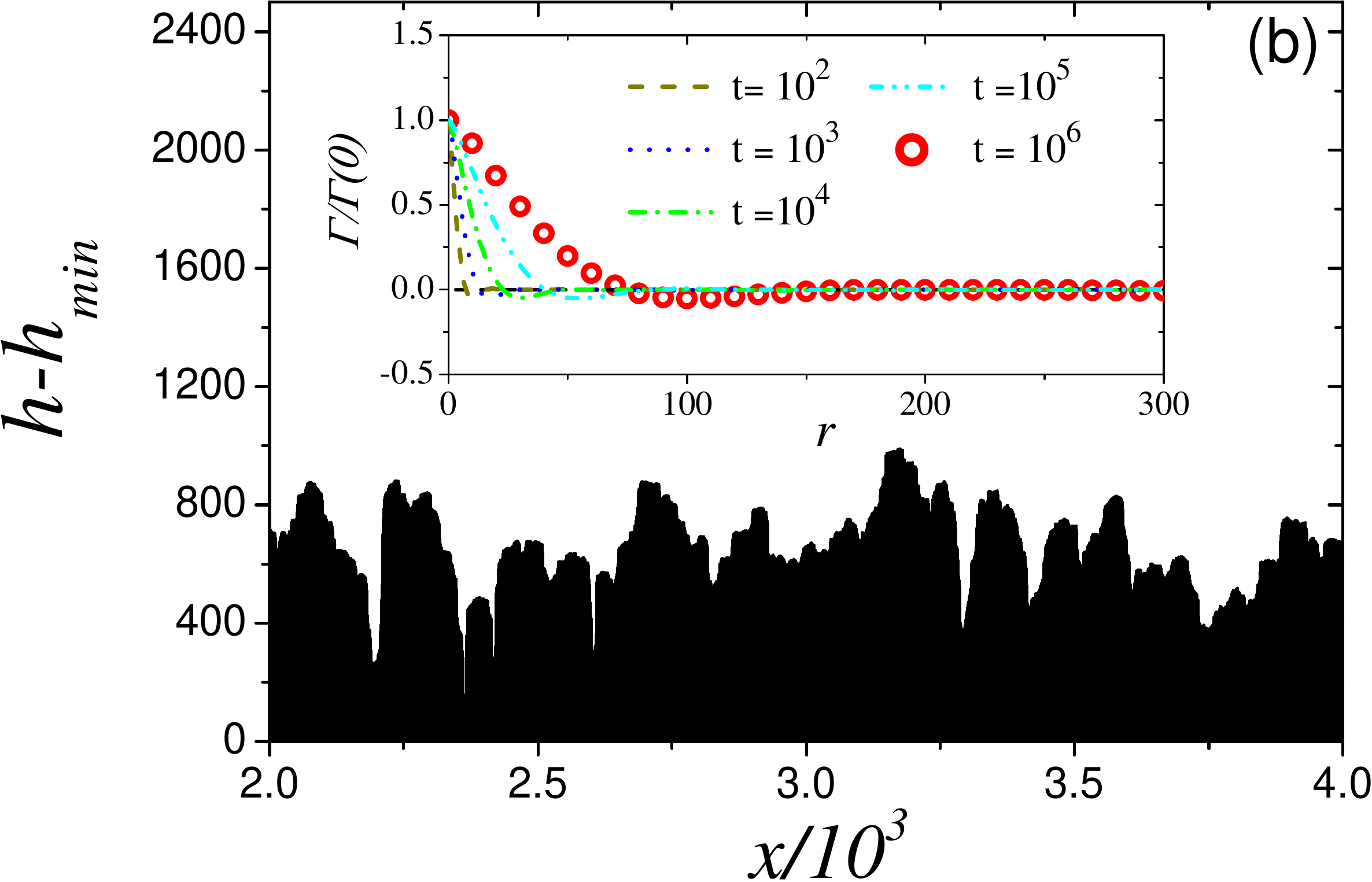}\\
	\includegraphics [width=0.4936\linewidth] {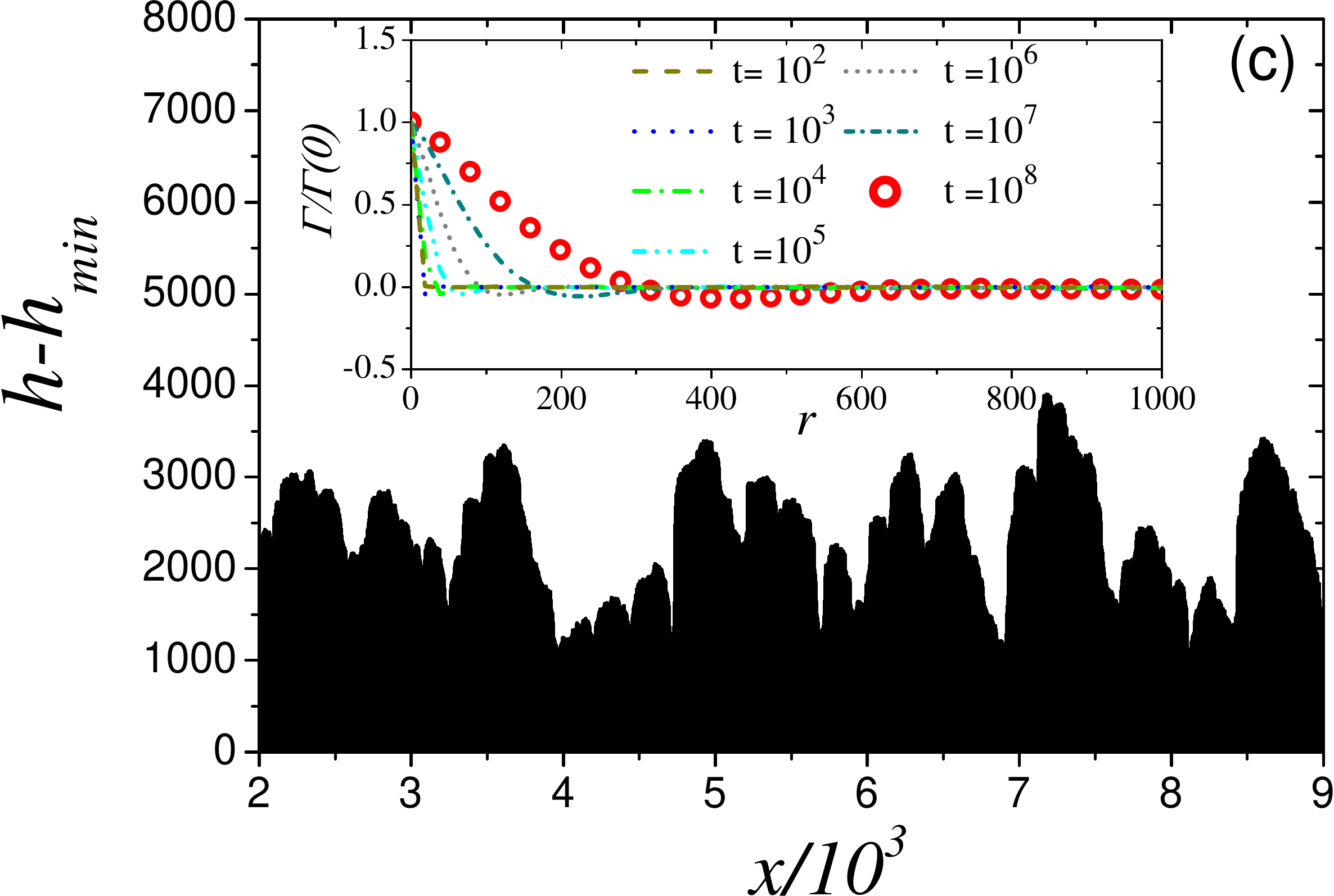}
	\includegraphics [width=0.4936\linewidth] {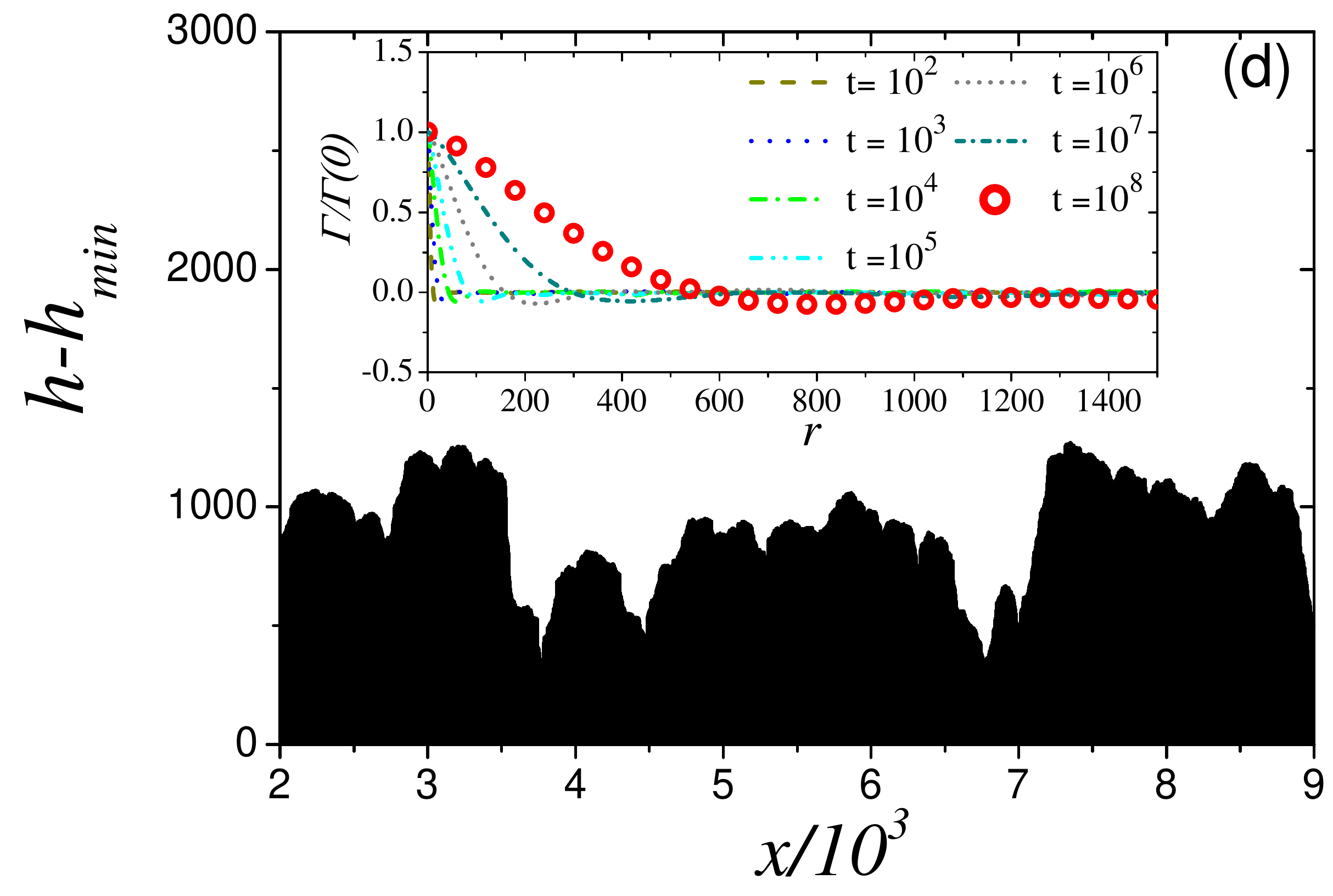}
	\caption{Profiles for (a) CRSOS,  (b) CV, and (c,d) DT models  at $t=10^{6}$
		(CRSOS and CV)  and  $t=10^{8}$ (DT). The simulations were performed on systems
		of size $L=2^{14}$, assuring that the dynamics is not in the stationary regime
		of roughness saturation for the analyzed times. Insets show the corresponding
		normalized correlation function at different times, averaged over $10^{3}$
		independent realizations. Analysis for the DT model using noise reduction with
		$M=4$ is shown in (d).
		} \label{Fig:Corr}
\end{figure*}

Figure~\ref{Fig:Corr} shows profiles for  CRSOS, CV and DT models for times
$t=10^6$ (CRSOS and CV) and $t=10^{8}$ (DT). One can see the presence of a
characteristic length (mounded structures) for the CV and DT cases and a
self-affine structure with less evident mounds for CRSOS model. As illustrated by the corresponding insets in Fig.~\ref{Fig:Corr}, the estimated characteristic lengths $\xi_0$ for CRSOS and CV correspond to $\xi_{0} \approx 433$ and $\xi_{0} \approx 74$, respectively. For DT, the estimated values without and with noise reduction are $\xi_{0} \approx 299$ and $\xi_{0}=564$, respectively. Here, it is possible to note a decrease of the global roughness as $M$ increases suppressing large hills and valleys in the surfaces. Concomitantly, an increase of the characteristic mound sizes is observed, which is reflected by an increasing of the correlation length.

\section{Optimal Detrended Fluctuation Method}
\label{sec:sorderdfa}

Let us start with the standard DFA method using a $n$th order polynomial to
detrend the surface~\cite{Peng1994}, called hereafter of DFA$_{n}$. The interface
fluctuation within a window $i$ of size $r$ in DFA$_n$ is defined as
\begin{equation}
\omega^{(n)}_i =
\lrangle{(\delta^{(n)})^2}_i^{1/2}
\label{Eq9}
\end{equation}
where
\begin{equation}
\delta^{(n)}=h(x)-G_i(x;A^{(0)}_i,A^{(1)}_i,\ldots,A^{(n)}_i),
\label{delta1}
\end{equation}
 $G_i$ is a $n$th order polynomial regression of the interface within the $i$th
 window with coefficients $A^{(0)}_i,A^{(1)}_i,\ldots,A^{(n)}_i$ obtained using
 least-square method~\cite{NR}. The local roughness yielded by the DFA$_n$
 method $\omega^{(n)}$ is given by the average over different windows and
 samples. In the standard local roughness analysis, that corresponds to DFA$_0$,
 the surfaces fluctuations are computed in relation to the average height such
 that $G_i=A^{(0)}_i=\lrangle{h}_i$.

In the ODFA method, the local roughness in the window $i$ of size $r$ is defined by Eq.~(\ref{Eq9}) with
\begin{equation}
\delta^{(n)}=\min_{x}\left[h(x)-G_i(x;A^{(0)}_i,A^{(1)}_i,\ldots,A^{(n)}_i)\right],
\label{deltamin}
\end{equation}
where $\min_x$ represents minimal distance from the surface point with height $h(x)$ to the polynomial $G_i$.

Differences between the exponents yielded by DFA and ODFA methods were
reported~\cite{Luis2017} in the kinetic roughening obtained for the deposition on
initially mounded substrates. The second order ODFA$_2$  method allows to
capture the expected universality class of the fluctuations at scales shorter
than the average mound length, whereas DFA$_2$ underestimates the
exponents~\cite{Luis2017}. In both cases, the main advantage of the second order
methods with respect to the first order ones are more extended regions of
scaling, represented by longer plateaus in  the effective roughness exponent,
\begin{equation}
\alpha_\mathrm{eff} \equiv \frac{d\left[\ln \omega^{(n)}\right]}{d\left[\ln r\right]},
\end{equation}
as function of the scale $r$.

\section{Scaling of the Local Surface Roughness}
\label{sec:Slocal}

\begin{figure*}[hbt!]
	\includegraphics [width=0.4936\linewidth]{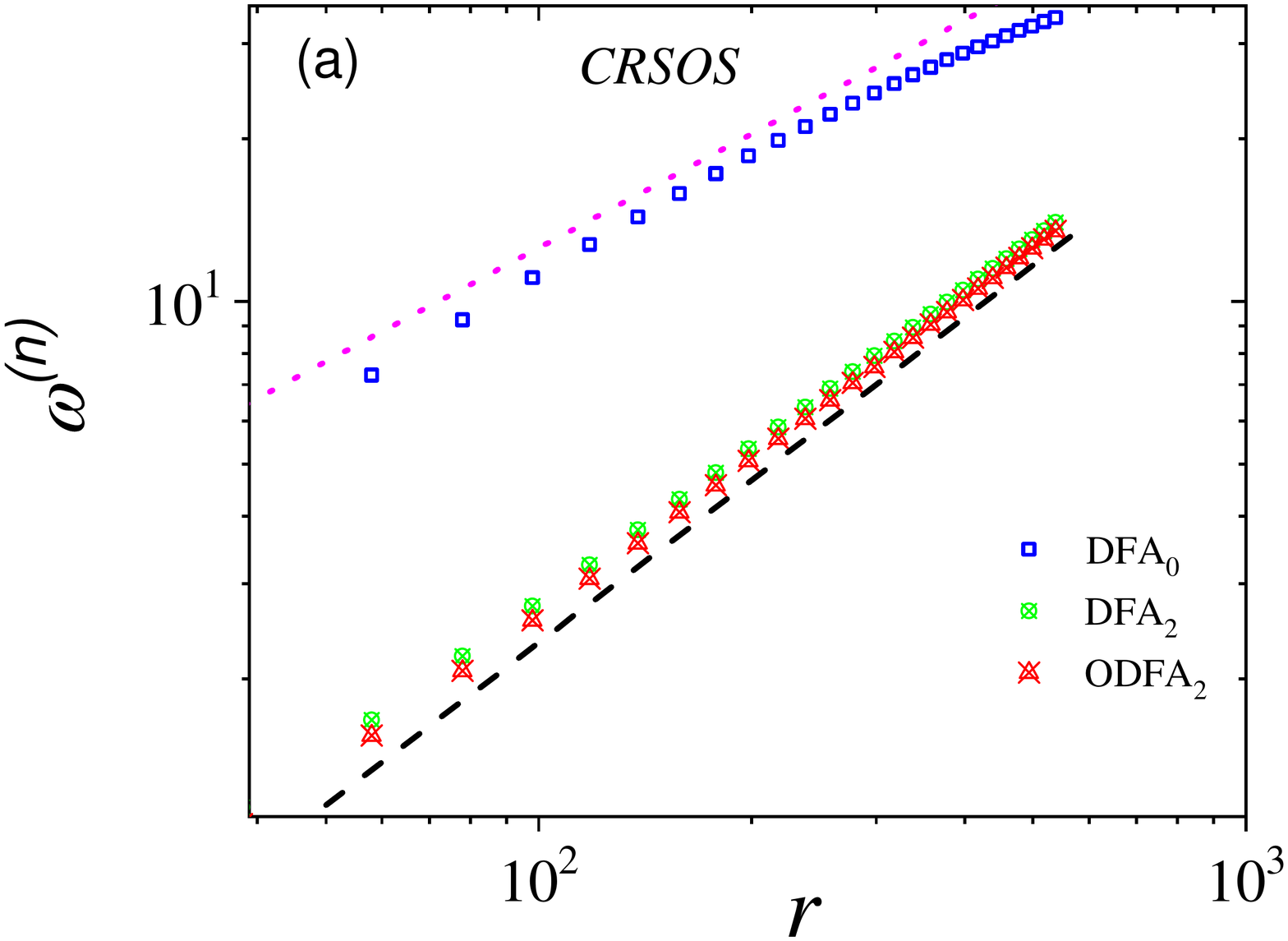}
	\includegraphics [width=0.4936\linewidth]{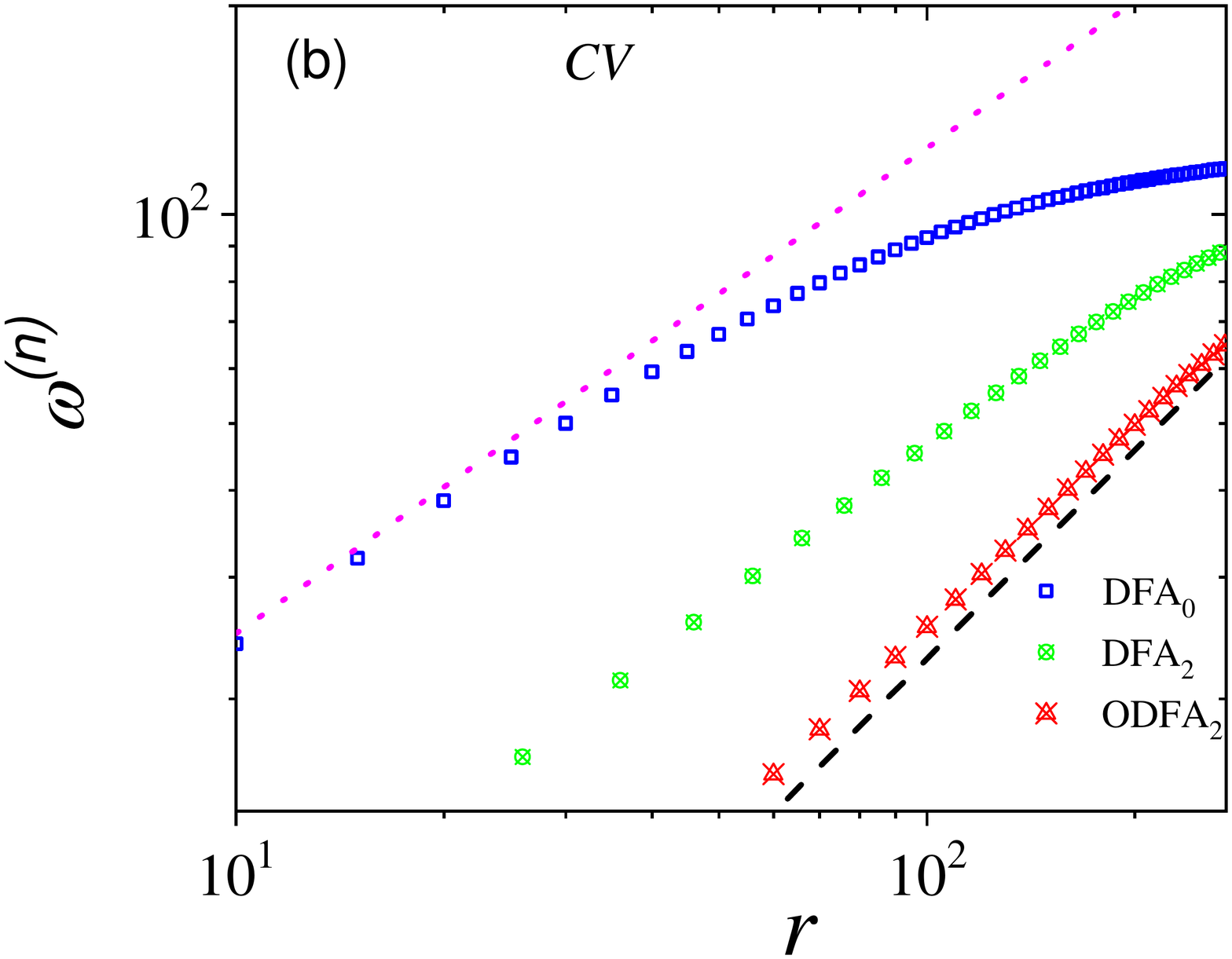}\\
	\includegraphics [width=0.4936\linewidth]{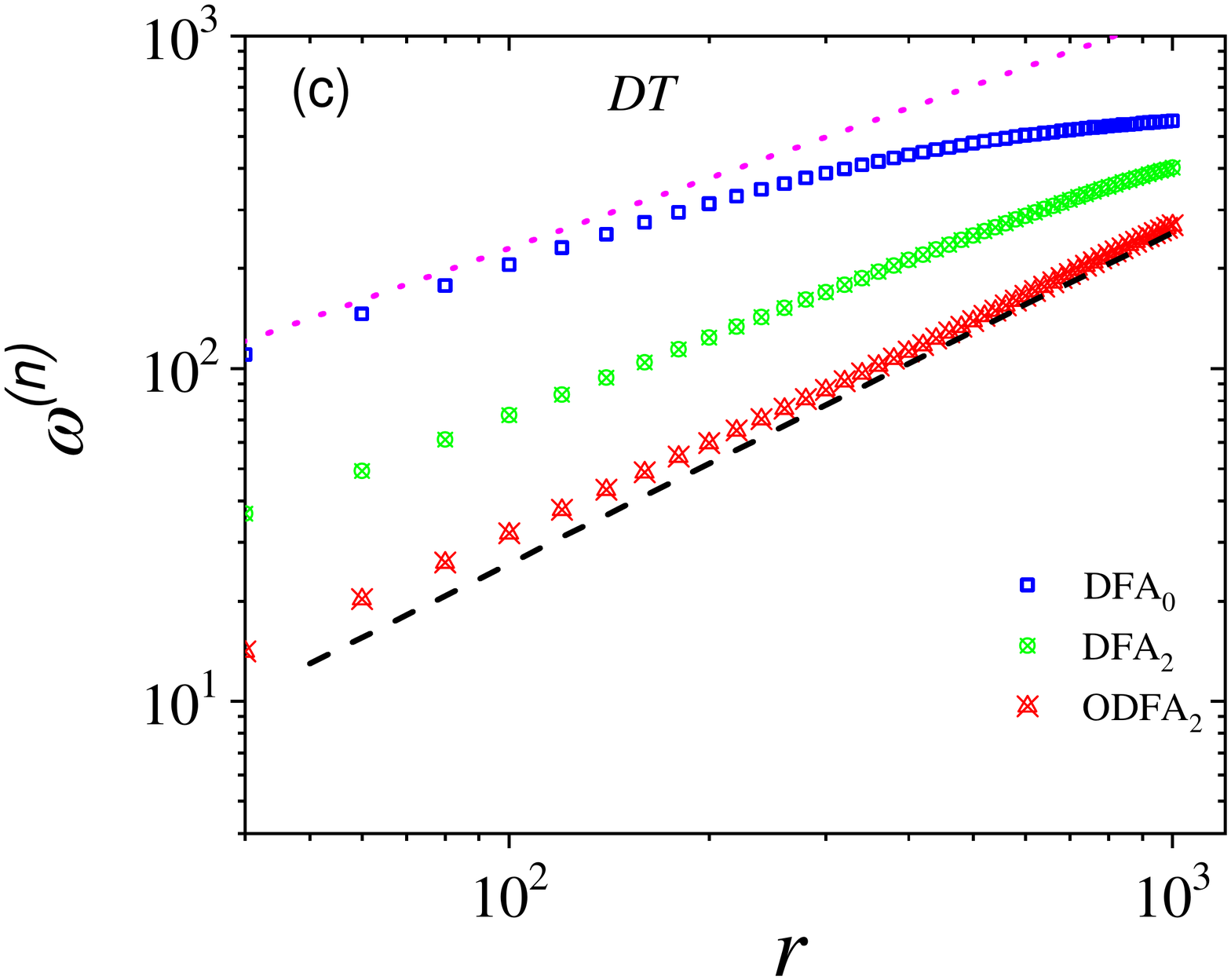}
	\includegraphics[width=0.4936\linewidth]{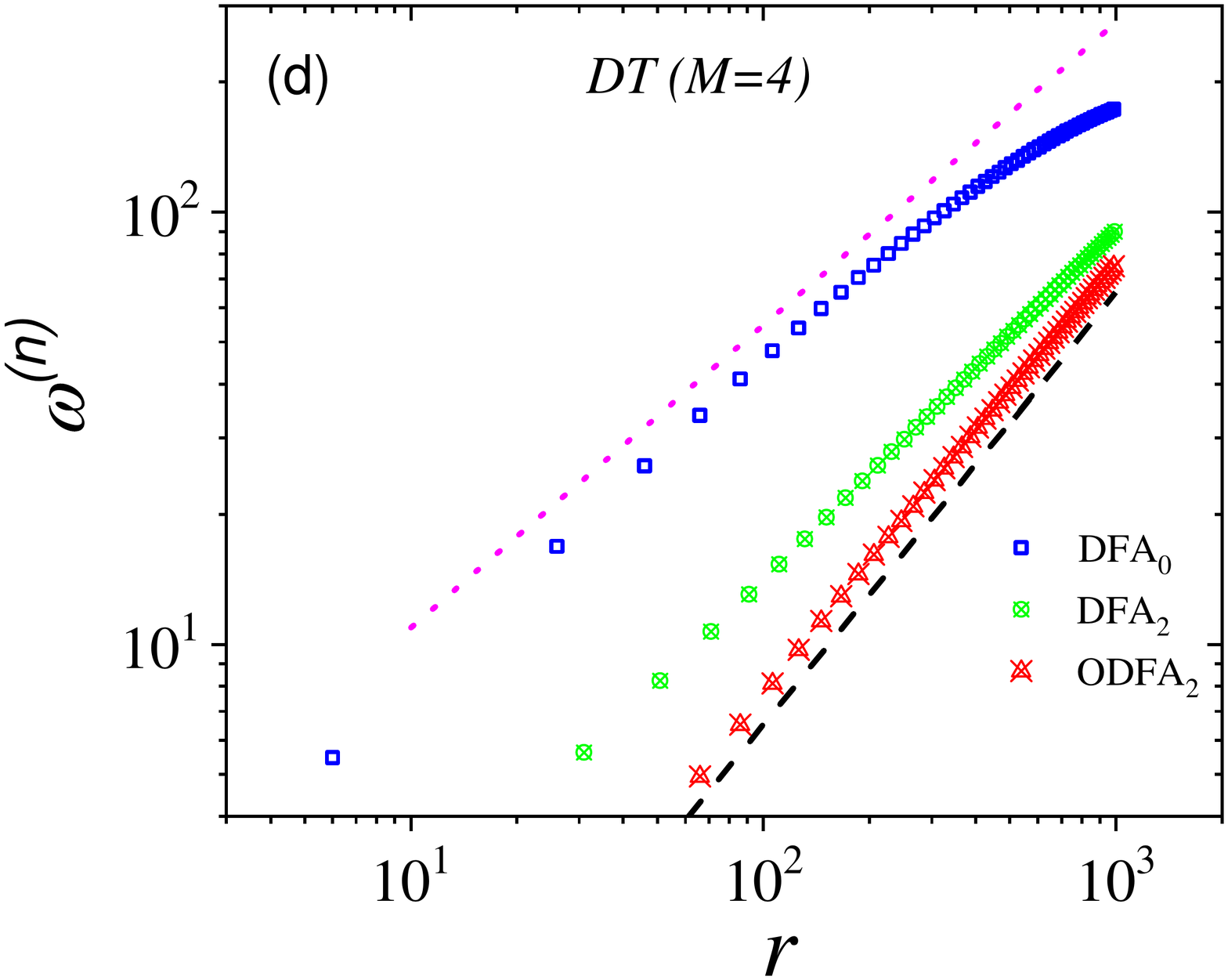}
	\caption{Local roughness as a function of the window size using different
		methods for (a) CRSOS model at t=$10^6$; (b) CV model at t=$10^{6}$
		and (c) DT model at $t=10^{8}$. The dotted and dashed lines
		have slopes $0.7$ and $1$, respectively. Averages were performed over up to
		$10^{3}$ independent realizations. The system size is $L=2^{14}$. (d) The same
		analysis for the DT model using noise reduction with $M=4$~\cite{Xun2015}.
		}
	\label{Fig:Omega}
\end{figure*}

Figure~\ref{Fig:Omega} shows the local roughness $\omega^{(n)}$ as a function of
the window size $r$ for CRSOS, CV and DT models. The analyses using DFA$_{0}$
indicate a local slope close to $0.7$ at small scales ($r\lesssim10^{2}$) for
all cases corroborating previous reports for models in the
VLDS universality class~\cite{Krug1994,Chame2004}. However, the slopes are close to $\alphaloc{}=1$,
predicted by the one-looping RG approach~\cite{Janssen1997}, when we consider the
scaling obtained from ODFA$_{2}$.

\begin{figure*}[hbt!]
	\includegraphics [width=0.4936\linewidth]{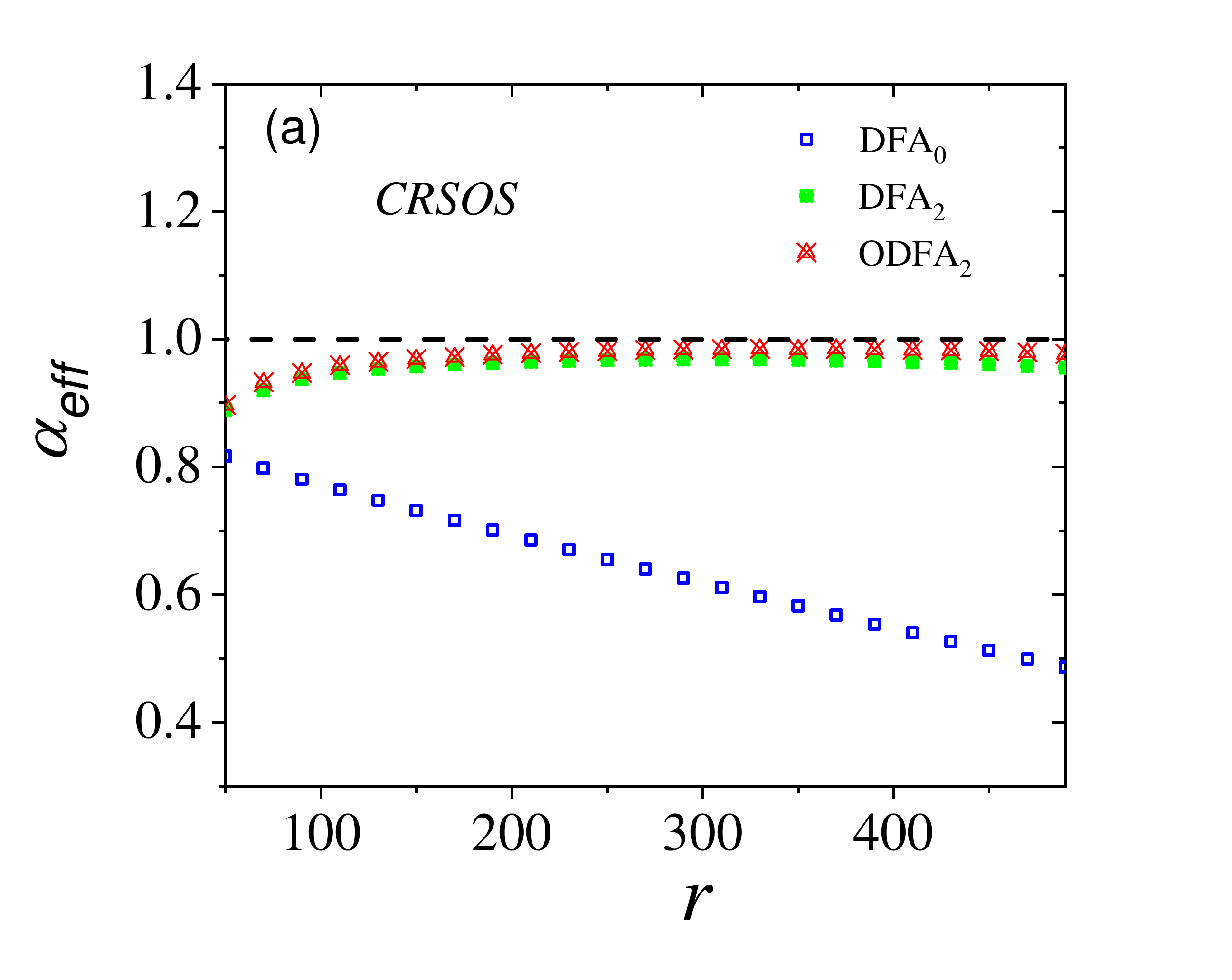}
	\includegraphics [width=0.4936\linewidth]{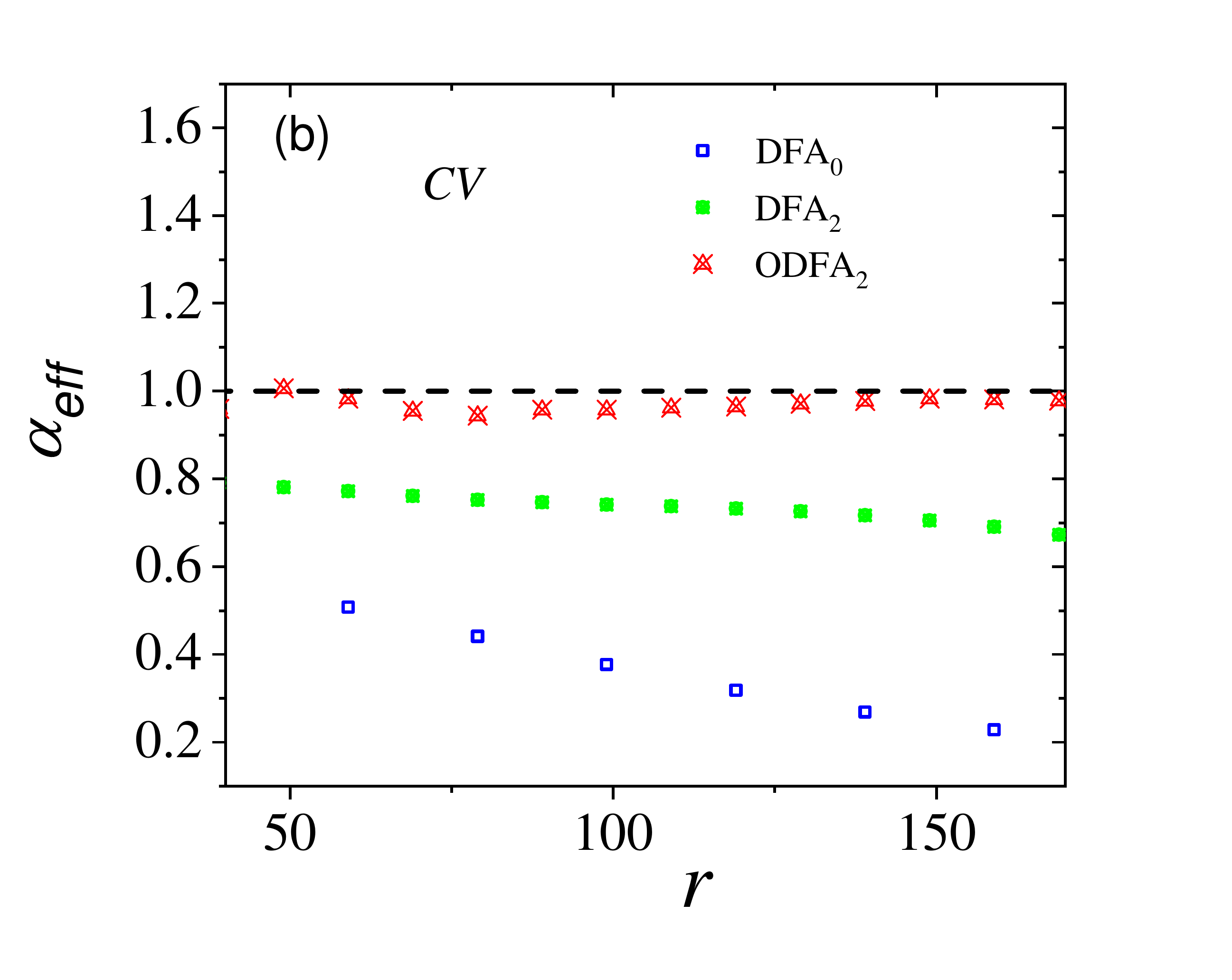}\\
	\includegraphics [width=0.4936\linewidth]{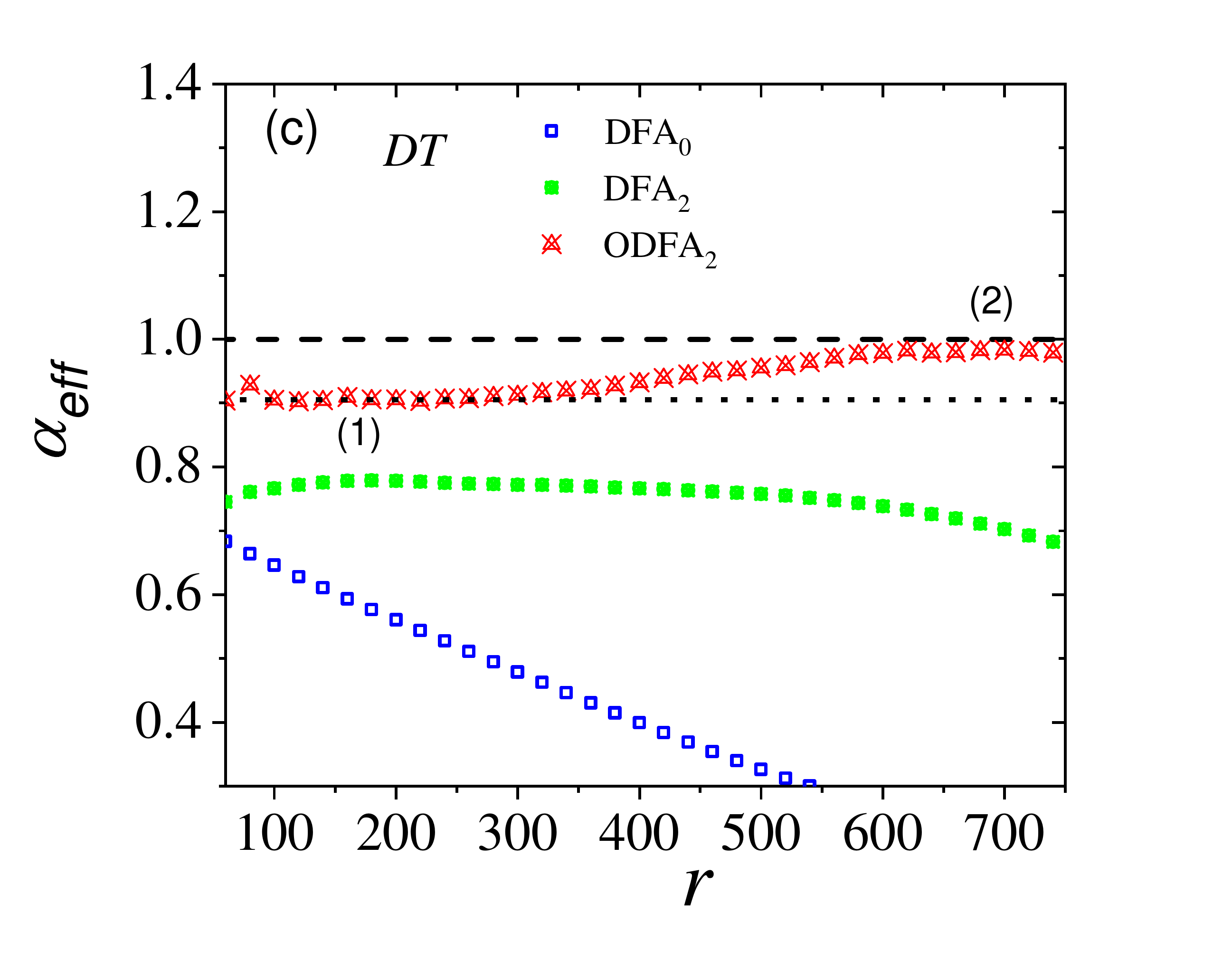}
	\includegraphics[width=0.4936\linewidth]{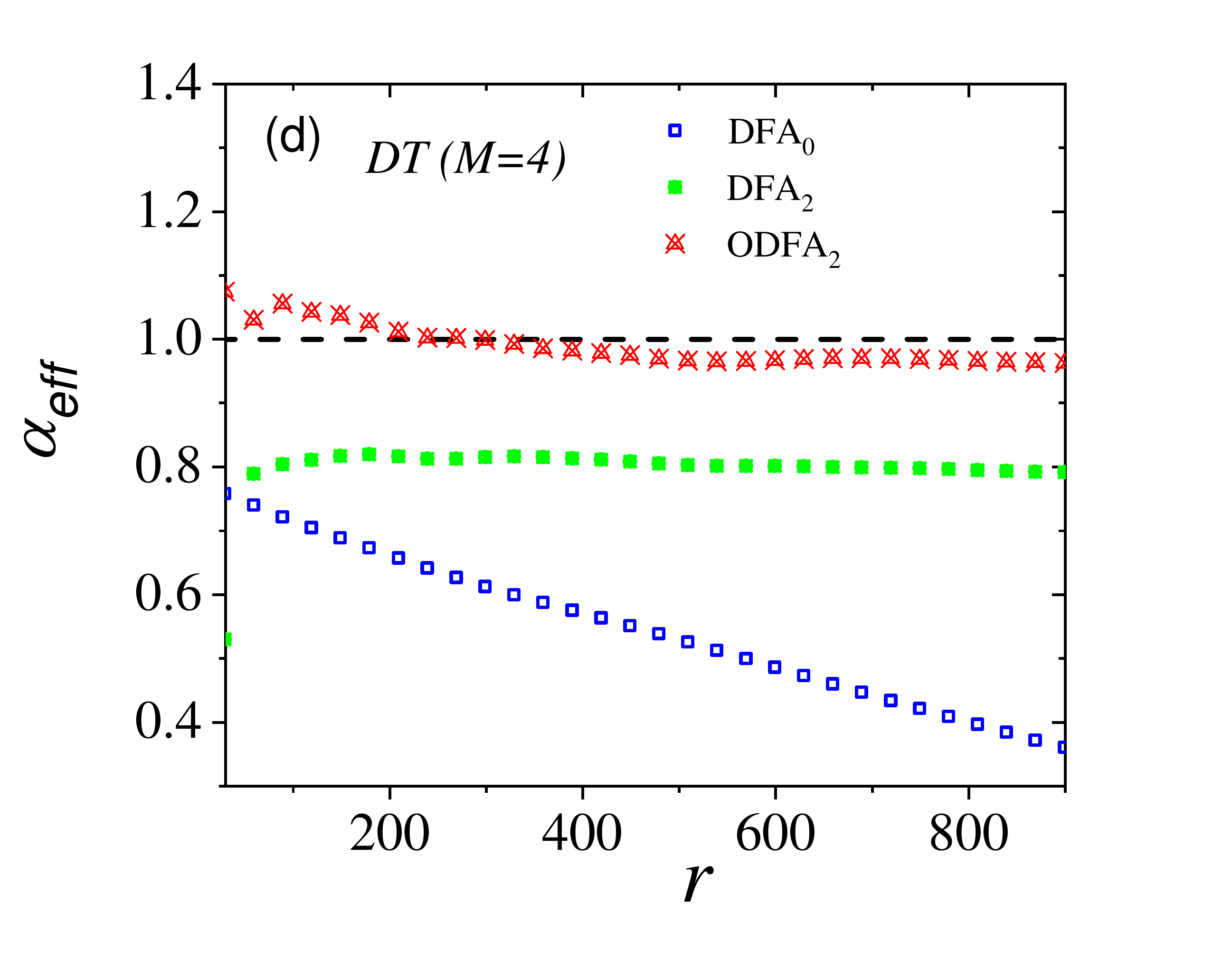}
	\caption{Effective local roughness exponent analysis
		with different methods for (a) CRSOS model at t=$10^6$; (b) CV model at
		t=$10^{6}$ and (c) DT model at $t=10^{8}$. Two scaling regions
		are observed in the DT model at shorter and larger scales indicated by (1) and (2).
		The horizontal dashed lines indicate the value of the nMBE roughness exponent
		predicted by one-looping RG in $d=1$ and the dotted line in (c)
		indicates the value $0.90$. (d) The same analysis for the DT model using noise
		reduction with $M=4$~\cite{Xun2015}.}
	\label{Fig:alpha_eff}
\end{figure*}

\begin{figure*}[hbt!]
	\includegraphics [width=0.48\linewidth] {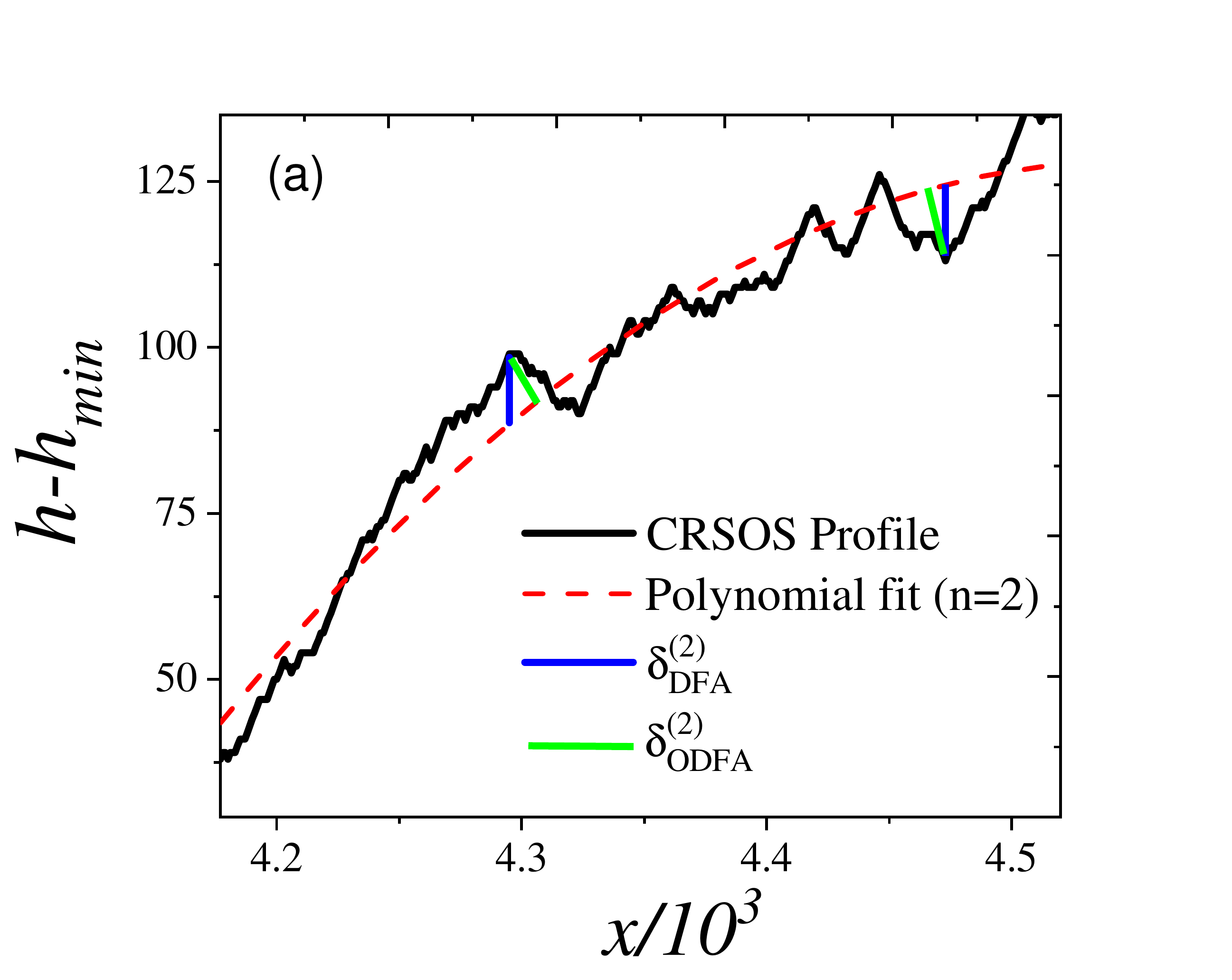}
	\includegraphics [width=0.4932\linewidth] {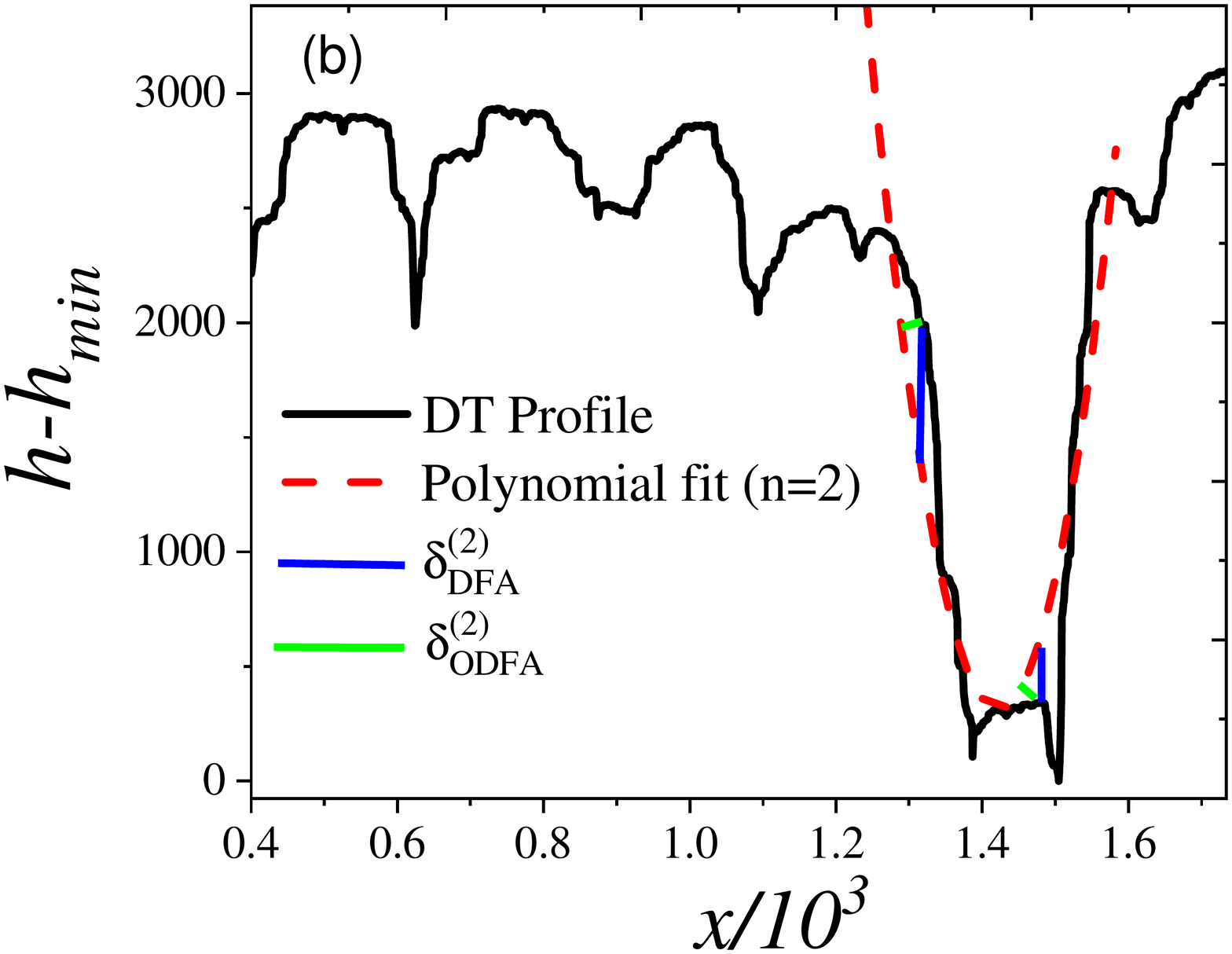}
	\includegraphics [width=0.4932\linewidth]  {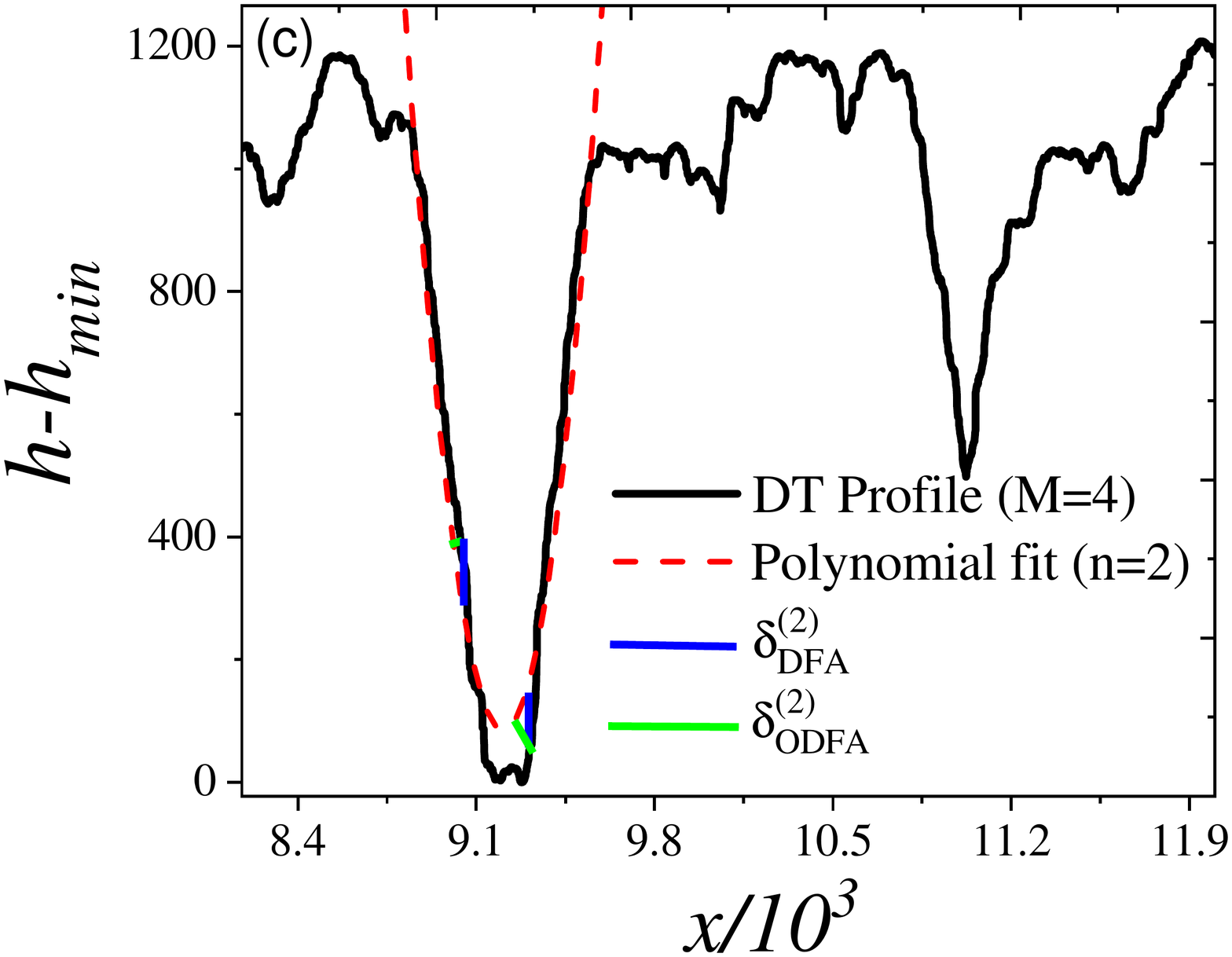}
	\caption{Sections of the profiles shown with the second order polynomial
		regressions (dashed lines) for (a) CRSOS and (b)-(c) DT models. DT model is analyzed (b) without and (c) with noise
		reduction using $M=4$. Selected points of the profile illustrates the
		differences between the $\delta^{(2)}$ calculated with DFA and ODFA
		methods.}
	\label{Fig:insight}
\end{figure*}

In the case of the CRSOS model, DFA$_{2}$ and ODFA$_{2}$ methods provide very
similar curves, confirmed in the local roughness exponent analysis of Figs.~\ref{Fig:Omega}(a) and
Fig.~\ref{Fig:alpha_eff}(a). This can be justified by the self-affine (fractal)
geometry exhibited by the profile, as observed in Fig.~\ref{Fig:Corr}(a), which implies in
negligible differences between the height fluctuations determined either by
Eq.~\eqref{delta1} or \eqref{deltamin}. This fact is illustrated in Fig.
\ref{Fig:insight}(a), in which a zoomed part of a CRSOS profile is shown with
the respective differences $\delta^{(2)}_\text{DFA}$ and
$\delta^{(2)}_\text{ODFA}$ for some selected points. We also verified that the
corresponding scaling of the second order methods are improved (the plateau
region of the effective exponent analysis is larger) if compared with their first order counterpart, but the exponent values are approximately the same
(results not shown). We determined the local roughness exponent of the CRSOS
model in the plateau $260 \lesssim r \lesssim 460$ shown in
Fig.~\ref{Fig:alpha_eff}(a) and found $\alphaloc{(CRSOS)}=0.983(1)$, which is
consistent with the claim of corrections in the one-loop RG analysis such that
$\alphaloc{}=1-\epsilon$~\cite{Janssen1997}. Our result suggests that the
corrections in the one-loop RG exponent are consistent with but not equal to that
reported in two-loop RG calculations \cite{Janssen1997}, as  previously
indicated elsewhere~\cite{AaraoReis2004b} for low dimensions. We stress that our
result for $\alphaloc{}$ obtained for the CRSOS model, in which weak corrections
to the scaling are expected, is slightly above (4\% of deviation) to the global
roughness exponent $\alpha=0.94(2)$ reported in Ref.~\cite{AaraoReis2004b},
corroborating that the asymptotic anomalous scaling does not occur for this
model.


Differences between ODFA$_{2}$ and DFA$_2$ methods are more evident for CV model
as can be seen in Figs. ~\ref{Fig:Omega}(b) and ~\ref{Fig:alpha_eff}(b). Again, the plateau is larger
considering ODFA$_{2}$ as compared with ODFA$_{1}$ (data not shown). However,
for ODFA$_{1}$ we obtained an effective roughness exponent $\alphaloc{(CV)}\approx 1.14(4)$ slightly larger than
unity at small
scales, a spurious value that can be explained as a consequence of the large local slope in approximately
columnar parts of the profile, as shown in Fig.~\ref{Fig:Corr}(b). Indeed, the linear
regression does not fit well the corresponding structures at small scales while
the quadratic one does. Using the range $50 \lesssim r \lesssim 170$, which corresponds to the plateau shown in Fig.~\ref{Fig:alpha_eff}(b), the exponent obtained with ODFA$_2$ is $\alphaloc{(CV)}=0.96(1)$, in consonance  with the normal scaling of the CV
model observed in two dimensions ~\cite{DeAssis2015b,Luis2017}.

Figure~\ref{Fig:Omega}(c) shows the local interface roughness for the DT model
without noise reduction, in which noticeable differences can be seen in
ODFA$_{2}$ and DFA$_2$ methods.  One can see in Fig.~\ref{Fig:insight}(b) that
the differences between the distances to the detrending curve  using ODFA and
DFA methods can be very large. A crossover between two scaling regimes is
observed. For scales smaller than the characteristic length ($0.23 \lesssim
r/\xi_{0} \lesssim0.90$), a plateau is observed for ODFA$_{2}$ case, where the
local roughness exponent was estimated as $\alphaloc{(DT)}=0.903(1)$, consistent
with those found in the case of high noise reduction $M\ge 64$~\cite{Xun2015}. At
larger scales ($2.00 \lesssim r/\xi_{0} \lesssim 2.51$), $\alphaloc{(DT)}=0.976(1)$ was observed. To the best of our knowledge, ODFA$_2$ method for DT model yields a
first evidence, without noise reduction
techniques~\cite{Punyindu1998,Xun2015,Disrattakit2016}, of a roughness exponent consistent
with the nMBE class (see Fig.~\ref{Fig:alpha_eff}(c)). Even though  noise reduction should not change the
universality class~\cite{Kertesz1988}, very high levels reduce a lot the
interface roughness, which becomes smoothed with a trending to provide an
exponent close to 1. So, we have also analyzed the DT model with a mild  noise
reduction ($M=4$). The interface roughness is reduced with respect to the original model but is still quite large, as can be seen in
Fig.~\ref{Fig:insight}(c). The local roughness analyses are shown in
Figs~\ref{Fig:Omega}(d) and \ref{Fig:alpha_eff}(d). With ODFA$_2$ we observed
$\alphaloc{(DT)}=0.967(2)$, which is much closer to VLDS class than the value
$\alphaloc{(DT)}=0.804(7)$ obtained with DFA$_2$. The latter is in good
agreement with the $\alphaloc{}$ exponent reported in Ref.~\cite{Xun2015} for a similar noise reduction parameter. This result could lead to a misinterpretation supporting anomalous scaling in DT model, since a
large global roughness exponent $\alpha\approx 1.2$ was also reported in Ref.~\cite{Xun2015} for this same range of $M$.
Our resuls with ODFA$_2$ strongly suggest the absence of the anomalous scaling for the
DT model too.

\section{Conclusions}
\label{sec:conclu}

The scaling properties of one-dimensional interfaces obtained with simulations
of lattice models belonging to nMBE universality class is an issue that has
attracted considerably
attention~\cite{Park2001,Chame2004,Leal2011a,To2018,Disrattakit2016,Xun2015,Luis2017}, given the outstanding importance of diffusion for applications in thin film
growth~\cite{Evans2006,Ohring2002}. 
Differently from the Kardar-Parisi-Zhang~\cite{Kardar1986} universality class that
have a plenty of lattices models described by its scaling
exponents~\cite{Barabasi1995,Krug1997}, the nMBE class, where diffusion is the
ruling mechanism on the surface growth, has only a few basic prototypes. Three basic examples are the
CRSOS~\cite{Kim1994,Kim1997}, CV~\cite{Clarke1987,Clarke1988} and DT~\cite{DasSarma1991} models, which are investigated in the present work. To date, only the first one has been supported with
irrefutable evidences that it belongs to the nMBE class. In the present work, we provide numerical analysis of the local roughness (Hurst) exponent~\cite{Barabasi1995,Peng1994} of interfaces generated with these models, using the recently proposed optimal detrended fluctuation analysis~\cite{Luis2017}, that is devised to investigate universality class in mounded structures. As in the two-dimensional analysis of mounded surfaces, the ODFA
method~\cite{Luis2017} outperforms the standard DFA in the determination of the
local roughness exponent $\alphaloc{}$, as can be seen in table~\ref{tabII},
where a summary of  the results  reported in this paper is presented. For all
investigated models, the roughness exponent were found within the interval
$[0.96,0.98]$. This rules out the existence of asymptotic anomalous roughening
sometimes claimed for these models~\cite{Xia2013} since these values are
consistent with the predictions of the two-loop renormalization group developed
by Janssen~\cite{Janssen1997}, where corrections in one-loop calculations of the
form $\alphaloc{(VLDS)}=1-\epsilon$ are expected in all dimensions. To the best
of our knowledge, this is the first evidence for the local roughness exponent
measured in the DT model that agrees with the nMBE class. The original DT model still possesses a small ambiguity in the value of $\alphaloc{}$: at short scales, the value $\alphaloc{(DT)}=0.903$ is close to the VLDS exponent $\alphaloc{(VLDS)}=1-\epsilon$, albeit still not negligibly below the VLDS value observed for intermediary scales, as shown in table~\ref{tabII}. However, a mild noise reduction is
sufficient to remove very strong corrections to the scaling and an accordance
with the nMBE exponent  is also found.
\begin{table}[hbt!]
	\centering
	\renewcommand{\arraystretch}{1.5}
	
	\caption{Values of the local roughness exponent calculated using the
		ODFA$_{2}$  and DFA$_{2}$ methods in the corresponding range of the ratio $r/\xi_{0}$ (see
		also Fig.~\ref{Fig:alpha_eff} for plateaus in the effective exponent analysis).
		The symbol $*$ means the absence of reliable scaling regimes
                                       in the corresponding range of $r/\xi_{0}$.}
	\begin{tabular*}{\linewidth}{c@{\extracolsep{\fill}}ccc}
		\hline\hline
		Model & $r/\xi_{0}$ & ODFA$_{2}$ &  DFA$_{2}$ \\
		\hline
		CRSOS & $[0.6, 1.06]$ & $0.983(1)$ & $0.966(2)$  \\
		CV & $[0.67, 2.3]$ & $0.96(1)$  & $0.73(3)$ \\
		DT [Region (1)] & $[0.23,0.90]$ & $0.903(1)$  & $0.772(6)$ \\
		DT [Region (2)] & $[2.00,2.51]$ & $0.976(1)$ & $*$\\
		DT [$M=4$] & $[0.88,1.58]$          & $0.967(2)$& $0.804(7)$ \\ \hline\hline
	\end{tabular*}
	
	\label{tabII}
\end{table}

Our findings constitute an important step for confirming the nMBE as a general
universality class.  Moreover, the scarcity of experimental evidences for nMBE
could be explained by the almost unavoidable presence of strong corrections to
the scaling due to limitations for growth times and resolution in scanning probe microscopes~\cite{Lechenault2010,Alves2016c}, which might be addressed using suitable
methods such as ODFA~\cite{Luis2017}. This method can be easily extended to
the analysis of self-affine objects not related to surface growth such as time series
modulated for seasonal changes~\cite{Vassoler2012}. Further enhancement of this method
may include adapting it for global detrending which will allow the characterization of other
features in interface growth such as properties of the underlying fluctuations in
height distributions~\cite{Alves2014,Carrasco2016,DeAssis2012}.

\begin{acknowledgments}
TAdA, SCF and RFSA thank the financial support of Conselho Nacional de Desenvolvimento Cient\'{i}fico e Tecnol\'{o}gico (CNPq). SCF acknowledges the support of Funda\c{c}\~{a}o de Pesquisa do Estado de Minas Gerais (FAPEMIG). This study was financed in part by the Coordena\c{c}\~{a}o de Aperfei\c{c}oamento de Pessoal de N\'{i}vel Superior - Brasil (CAPES) - Finance Code 001. \
\end{acknowledgments}


%

\end{document}